# "It depends on where AI is used": Players' attitude patterns and evaluative logics toward different AI applications in digital games

Ting-Chen Hsu[1[0009-0009-7656-2492]], Jiangxu Lin[1[0009-0000-5619-7824]], Wenran Chen[1[0009-0009-7370-4299]], Fei Qin[2[0009-0006-2950-0641]] and Zheyuan Zhang[1[0009-0004-3773-3221]]

[1] School of Animation and Digital Arts, Communication University of China, Beijing 100024, China
[2] School of Information Engineering, Lanzhou City University, Lanzhou 730070, China
tingchenhsu.ac@gmail.com, zhangzheyuan04@163.com

**Abstract.** As AI becomes increasingly embedded in digital games, players' attitudes depend not only on whether AI is used, but also on where and how it intervenes in gameplay. This study examines players' evaluative patterns toward eight AI application contexts, including intelligent NPCs, emergent narrative, dynamic balancing, recommendation systems, review and governance, art asset generation, co-creation gameplay, and gameplay evolution. Based on 1,856 valid open-ended responses from 310 questionnaires, we conducted thematic analysis to identify reasons for acceptance, rejection, and conditional acceptance. Results show that players welcomed AI when it enhanced immersion, personalization, novelty, efficiency, or convenience, but resisted it when it threatened creativity, emotional authenticity, autonomy, fairness, system stability, authorship, or accountability. We further identify six evaluative logics: experiential enrichment, instrumental efficiency, system reliability, agency and control, authorship and compliance, and human oversight. These preliminary findings highlight the context-sensitive nature of AI acceptance in digital games.

## 1 Introduction

In recent years, AI—especially generative AI—has been increasingly integrated into digital games, such as non-player character (NPC) interaction [1], content generation [2], and player co-creation [3]. Yet "AI in games" is not a single object of player evaluation. Players may value AI when it expands immersion and sense of presence, but reject it when it threatens sense of control, system stability, human creativity or fairness [1,4,5]. Therefore, understanding players' attitudes toward AI requires asking not only whether AI is used in games, but also where, how, and for what purpose it is used within the broader game ecosystem.

Prior work has examined player responses to human and AI-generated content [6] and to specific AI-enabled systems such as AI-driven characters, procedural narrative, and AI-supported design tools [7,8,9,10]. These studies show that player evaluations of AI are shaped not only by technical performance, but also by how players interpret AI's relationship to human designers, game systems, and their own play experience. However, existing research remains limited in three respects. First, it often focuses on individual systems or prototypes, making it difficult to distinguish attitudes toward AI in general from concerns about specific application contexts. Second, the range of AI applications studied remains narrower than the expanding use of AI in games, which now includes more aspects such as emergent gameplay and content co-generation. Third, less is known about the evaluative logics that explain why similar AI technologies may be welcomed in some contexts but resisted in others.

To address these gaps, this study investigates players' attitudes toward eight common AI application contexts in and around digital games. We conducted an open-ended survey (N=345, approximately 2,760 open-text responses were generated) to evaluate in detail the reasons why players accept or reject these different AI applications. Specifically, this study focuses on:

RQ1: What reasons do players give for accepting or rejecting AI in each application context?

RQ2: What commonalities and differences characterize these attitudes across AI applications?

RQ3: How can these reasons be abstracted into higher-level evaluative logics that explain variation in player attitudes?

This study makes three contributions. Theoretically, it develops a context-sensitive account of player acceptance and rejection by identifying six evaluative logics across eight AI application contexts. Empirically, it shows how concerns such as authenticity, fairness, stability, agency, and oversight are activated differently depending on where AI is used. Practically, it offers design implications for deciding where AI may enhance gameplay, where it may threaten player values, and where human oversight is needed.

# 2 Methods

## 2.1 Study design and questionnaire

This study used an open-ended online survey to examine how players evaluate different uses of AI in digital games. Participants were recruited through recruitment posts distributed on Chinese social media platforms. Participation was anonymous, and informed consent was obtained before respondents entered the questionnaire.

The questionnaire (Appendix A) collected participants' demographic and gaming background information, including age, gender, and weekly gaming time. Participants were then asked to explain their opinions for various AI applications across eight application contexts: driving intelligent NPCs, emergent narrative and task generation, dynamic balance adjustment, personalized matching and recommendation, report review and community governance, generating art assets, AI-supported co-creation gameplay, and dynamic evolution and emergence of gameplay. Each context was presented as a separate open-ended prompt.

## 2.2 Coding strategy

The responses were analyzed using codebook-based thematic analysis. After data preprocessing, two researchers independently pre-coded a random 20% sample of the open-text responses using a bottom-up approach to identify players' reasons across the eight contexts. Inter-rater reliability was assessed using Cohen's Kappa. Based on the pre-coding results and discussion, the research team developed a formal codebook containing code names, definitions, and examples (Appendix B). The same two researchers then coded the remaining 80% of the data in parallel using the finalized codebook. Because a single response could contain multiple reasons, multiple thematic codes could be assigned to the same text. Attitude polarity and thematic content were treated as two related but distinct coding dimensions. Each response was first classified as supportive, opposing, or conditional to its explicit evaluative position toward the AI application. The reasons underlying that position were then coded using the context-specific codebook. Thus, similar concerns could be compared across responses while their evaluative direction was retained. Disagreements were reviewed by a third senior researcher, who made final decisions to reach full consensus.

### 2.3 Data analysis

The coded data were analyzed in three steps. First, we constructed within-context profiles by examining the distribution and co-occurrence of thematic codes across supportive, opposing, and few conditional responses. Second, we compared themes across contexts using a context-by-theme matrix to examine shared concerns and context-specific differences; descriptive counts were used only as supplementary indicators of theme prevalence. Third, we conducted a second-cycle interpretive synthesis to abstract concrete themes into higher-level evaluative logics explaining how players' judgments varied across different uses of AI in games.

## 3 Results

### 3.1 Overview of data preprocessing and coding outcomes

A total of 345 questionnaires were collected, generating 2,760 open-ended responses. After excluding 35 invalid questionnaires with anomalous completion times or blank open-ended sections, 310 questionnaires and 2,480 responses entered data cleaning. A further 624 invalid individual responses (blank or only stating acceptance or rejection) were removed, leaving 1,856 valid open-text responses for thematic analysis. In the final sample, participants were aged 17-34 years (M=21.79), with 116 males, 187 females, and 7 non-binary. The professional backgrounds of the participants include natural sciences (8.71%), engineering and technology (30.65%), social sciences (16.13%), humanities and arts (39.03%), or other fields (5.48%).

To establish the coding scheme, two researchers independently pre-coded 20% of the cleaned data. Inter-coder reliability reached substantial agreement, with Cohen's Kappa of $\kappa$=0.71. Based on this process, a codebook was developed, containing approximately 7 to 12 themes for each application context. Across the 1,856 valid responses, 2,562 coded segments were identified, with an average of 1.38 codes per response. In total, 584 responses, or 31.47%, received more than one code.

### 3.2 Context-specific attitude reasons

Table 1 and Table 2 summarize players' reasons for accepting or rejecting AI across the eight application contexts. Overall, players' evaluations varied substantially by context rather than reflecting a general attitude toward AI in games.

For AI-driven NPCs, acceptance was mainly associated with immersion, presence, personalization, freedom, and reduced development cost, while resistance focused on illogical behavior, lack of emotion, system instability, and emotional burden. For emergent narrative and task generation, players valued personalization, richness, replayability, immersion, and longer gameplay lifespan, but worried about low creativity, chaotic narratives, low-quality output, and unstable systems. Some responses also noted that constantly changing narratives could either weaken shared community discussion or generate new discussion topics.

**Table 1.** Main supporting reasons across AI application contexts.

| AI application contexts | Supporting themes |
|---|---|
| Driving intelligent NPCs | High immersion/presence (64.7%); High personalization/richness (35.7%); High freedom (14.1%); Low development cost (1.7%); High logicality (0.4%) |
| Emergent narrative and task generation | High personalization/richness (37.6%); High novelty (36.7%); High replayability (17.1%); High immersion (14.7%); Longer gaming lifespan (7.3%) |
| Dynamic balance adjustment | High personalization/universality (50.4%); Reducing frustration (34.5%); Maintaining challenge (22.0%); High immersion (19.4%); Lower balancing costs (2.2%) |
| Personalized matching and recommendation systems | Higher gaming efficiency (36.0%); Catering to diverse preferences (36.0%); Enhanced interactive experience (34.1%) |
| Report review and community governance | High efficiency (57.6%); High fairness/objectivity (7.1%) |
| Generating art assets | Benefiting developers (35.2%); High completion (12.3%); High aesthetics (11.4%) |
| AI-supported co-creation gameplay | Meeting diverse demands (43.6%); High immersion (20.3%); High freedom/openness (18.1%); Lowering creation threshold (16.7%); High novelty/richness (11.0%); Improving creation efficiency (9.7%); Meeting sense of achievement (7.5%) |
| Dynamic evolution and emergence of gameplay | High novelty/richness (59.6%); Enhancing immersion/exploratory desire (24.5%); High gaming lifespan (21.1%); Meeting diverse demands (7.6%); Reducing developer workload (3.4%) |

Dynamic balance adjustment was supported when it reduced frustration, maintained challenge, or lowered balancing costs, but resisted when it disrupted rhythm, impaired autonomy, weakened achievement, or produced unstable balancing. Personalized matching and recommendation systems were valued for matching efficiency, preference diversity, and improved interaction, while concerns centered on monotony, inaccurate matching, reduced autonomy, and privacy risks. In report review and community governance, players recognized AI's potential for efficiency, fairness, and objectivity, but questioned its ability to avoid misjudgment, understand context, handle bias, and provide emotional feedback; many therefore framed this context as requiring human-AI collaboration.

For AI-generated art assets, acceptance was linked to developer benefits, content completion, and visual aesthetics, whereas resistance involved low innovation, weak authenticity or emotionality, copyright risks, limited controllability, and a perfunctory feeling. Some players supported AI when positioned as a human assistant. AI-supported co-creation was more positively associated with diverse demands, freedom, openness, lower creative thresholds, novelty, and creation efficiency, although concerns about output quality, authorship, and compliance remained. Finally, dynamic emergence of gameplay elements was valued for novelty, richness, immersion, exploration, longer gameplay lifespan, and reduced developer workload, but resisted when it disrupted balance, reduced stability or maturity, or increased cognitive load.

Taken together, acceptance was often tied to immersion, personalization, efficiency, novelty, and reduced workload, while resistance was associated with instability, low-quality output, re-duced autonomy, authorship or compliance risks, and the need for human oversight. These context-level differences provide the basis for the cross-context comparison and higher-level evaluative logics reported below.

**Table 2.** Main opposing reasons across AI application contexts.

| AI application contexts | Opposing themes |
|---|---|
| Driving intelligent NPCs | Out of character (4.6%); Low logicality (4.1%); Lack of emotions (3.7%); Low system stability (2.5%); High emotional burden (0.8%) |
| Emergent narrative and task generation | Lack of creativity/emotion (9.8%); Chaotic/low-quality narrative (9.0%); Low system stability (6.5%) |
| Dynamic balance adjustment | Unstable balancing (10.3%); Damaging game rhythm (6.0%); Impaired autonomy (6.5%); Lower sense of achievement (1.7%) |
| Personalized matching and recommendation systems | Prone to monotony (7.9%); Inaccurate matching (6.5%); Impaired autonomy (5.1%); Privacy concerns (2.3%) |
| Report review and community governance | Missed/misjudgment concerns (25.9%); Lack of semantic/emotional understanding (9.8%); Bias/fairness concerns (4.0%); Lack of emotional feedback (1.8%) |
| Generating art assets | Low innovation/creativity (15.3%); Low completion (15.3%); Lack of authenticity/emotionality (12.7%); Lack of creator-empathy (10.6%); Low aesthetics (7.2%); Copyright risk (3.4%); Low controllability (3.0%); Perfunctory feeling (1.7%) |
| AI-supported co-creation gameplay | Low quality (15.9%); Lack of authorship (4.4%); Copyright/compliance concerns (2.2%) |
| Dynamic evolution and emergence of gameplay | Disrupting balance/system stability (14.3%); Low quality (7.6%); Increasing cognitive load (5.9%) |

## 3.3 Shared and cross-context attitude patterns

To compare the eight contexts more systematically, the context-specific themes were grouped into 14 cross-context thematic clusters, as shown in Figure 1. Some clusters appeared across multiple AI applications, while others were concentrated in specific contexts.

Several supportive clusters recurred across contexts. Experience enhancement was most prominent in intelligent NPCs and also appeared in narrative generation, co-creation, and dynamic gameplay emergence. Personalization enhancement was especially salient in intelligent NPCs, narrative generation, dynamic balance adjustment, and recommendation systems. Convenience enhancement appeared mainly in system-facing and production-facing contexts, including recommendation systems, report review, art asset generation, and co-creation. Novelty and retention enhancement was most visible in narrative generation, dynamic balance adjustment, and dynamic gameplay emergence, suggesting that players accepted AI when it was expected to make gameplay feel fresher, richer, or more replayable.

Resistance clusters also showed clear contextual differences. Stability risk was concentrated in applications that directly intervene in gameplay systems, especially dynamic balance adjustment and dynamic gameplay emergence. Lower quality or performance appeared where AI output quality was highly visible, such as narrative generation, report review, art asset generation, and recommendation systems. Experience or emotional dissatisfaction was especially relevant to intelligent NPCs and art generation, where players expected emotional expression, human-like interaction, or creator sensitivity. Lack of creativity or novelty was mainly associated with narrative generation and AI-generated art assets, indicating that players did not evaluate all AI-generated content uniformly.

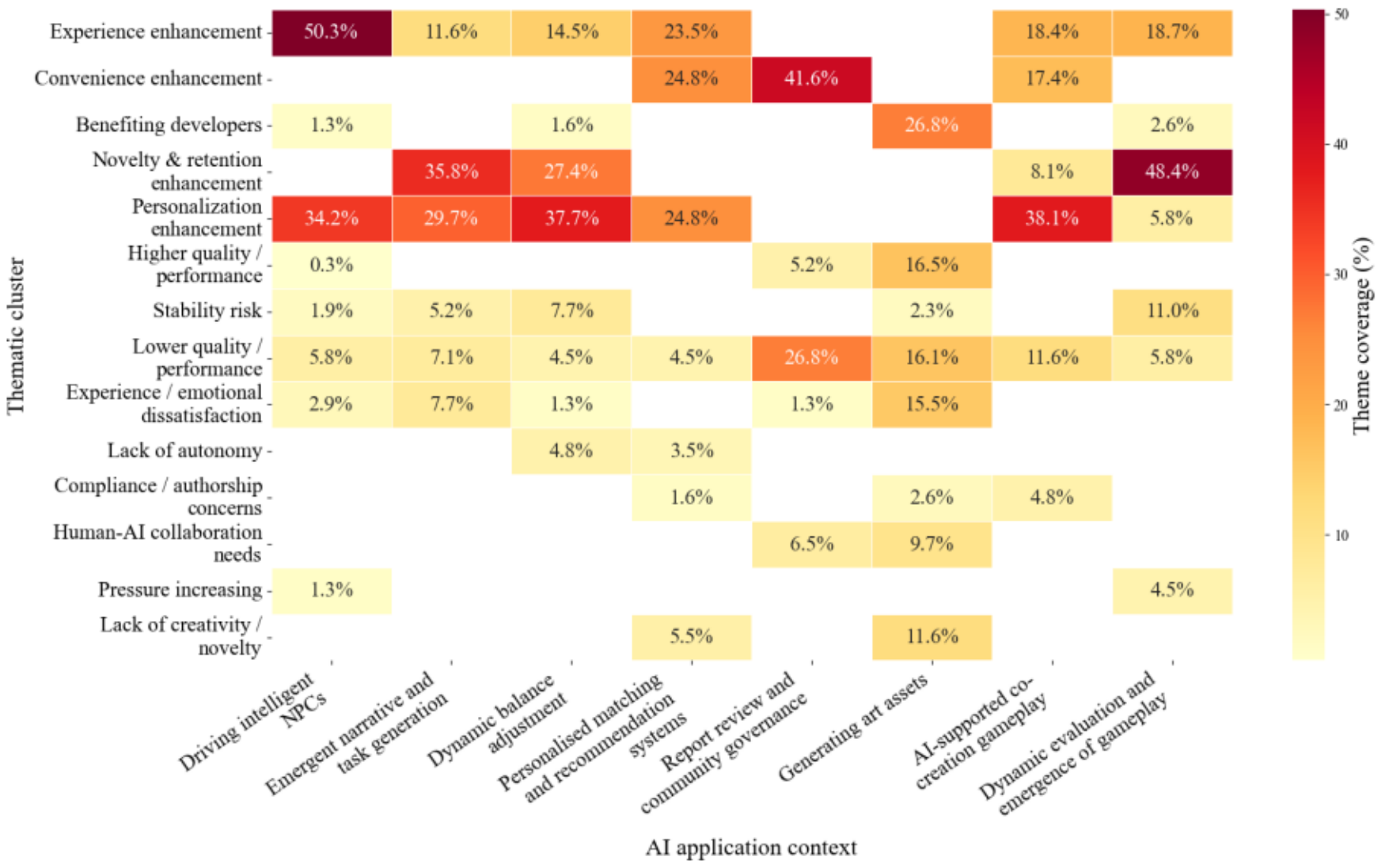

**Fig. 1.** Distribution of thematic clusters across AI application contexts

Authorship and compliance concerns were most prominent in AI-generated art assets and AI-supported co-creation, where players raised issues of copyright, ownership, and creative legitimacy. Lack of autonomy was especially associated with personalized matching and dynamic balance adjustment, reflecting concerns that AI might make decisions on behalf of players. Human-AI collaboration needs appeared most clearly in report review and AI-generated assets, suggesting that some players did not reject AI itself but preferred it to remain supervised, reviewed, or complemented by human judgment.

### 3.4 From reasons to evaluative logics

The thematic clusters were further synthesized into six higher-level evaluative logics, as shown in Table 3. These logics explain why players' attitudes varied across contexts: players were not evaluating AI as a single technology, but judging the type of intervention it performed within the game ecosystem.

First, the experiential enrichment logic concerns whether AI makes gameplay more immersive, personalized, fresh, or enjoyable. This logic was central to intelligent NPCs, narrative generation, co-creation, and dynamic gameplay emergence. Second, the instrumental efficiency logic concerns whether AI improves production, moderation, matching, or content-generation efficiency. It appeared strongly in recommendation systems, report review, and art asset generation, where AI was often treated as a practical tool.

Third, the system reliability logic concerns whether AI maintains or disrupts stability, rhythm, balance, and coherence. It was especially important in dynamic balance adjustment and dynamic emergence of gameplay elements. Fourth, the agency and control logic concerns whether AI reduces players' sense of choice, autonomy, or achievement, and was particularly relevant to matching systems and balance adjustment.

Fifth, the authorship and compliance logic concerns ownership, copyright, privacy, and creative legitimacy. It was most salient in AI-generated art assets and AI-supported co-creation. Sixth, the human oversight logic concerns whether AI should operate autonomously or under human supervision. This logic was especially visible in report review, governance, and asset generation, where players accepted AI more readily when it assisted rather than replaced human judgment.

Figure 2 integrates these findings by linking the eight application contexts to broader sites of AI intervention and to the six evaluative logics. The model summarizes how players connected different AI applications with experiential, instrumental, systemic, agential, authorial, and governance-related concerns.

**Table 3.** Higher-level evaluative logics underlying players' attitudes toward AI applications.

| High-level evaluative logic | Core judgment problem |
|---|---|
| Experiential enrichment logic | Does AI make the gaming experience more immersive, personalized, fresh, or enjoyable? |
| Instrumental efficiency logic | Does AI play a role as a tool to improve efficiency and quality, reduce costs, and enhance production or gaming convenience? |
| System reliability logic | Will AI affect system stability, game rhythm, balance, or lead to more chaos and uncertainty? |
| Agency and control logic | Does AI weaken player autonomy, make players feel restricted, intervened, or bring additional pressure to them? |
| Authorship and compliance logic | Does AI bring about authorship crisis, compliance risk, privacy infringement, or copyright issues? |
| Human oversight logic | Does AI need to be supervised, reviewed, led, or operated in collaboration with humans? |

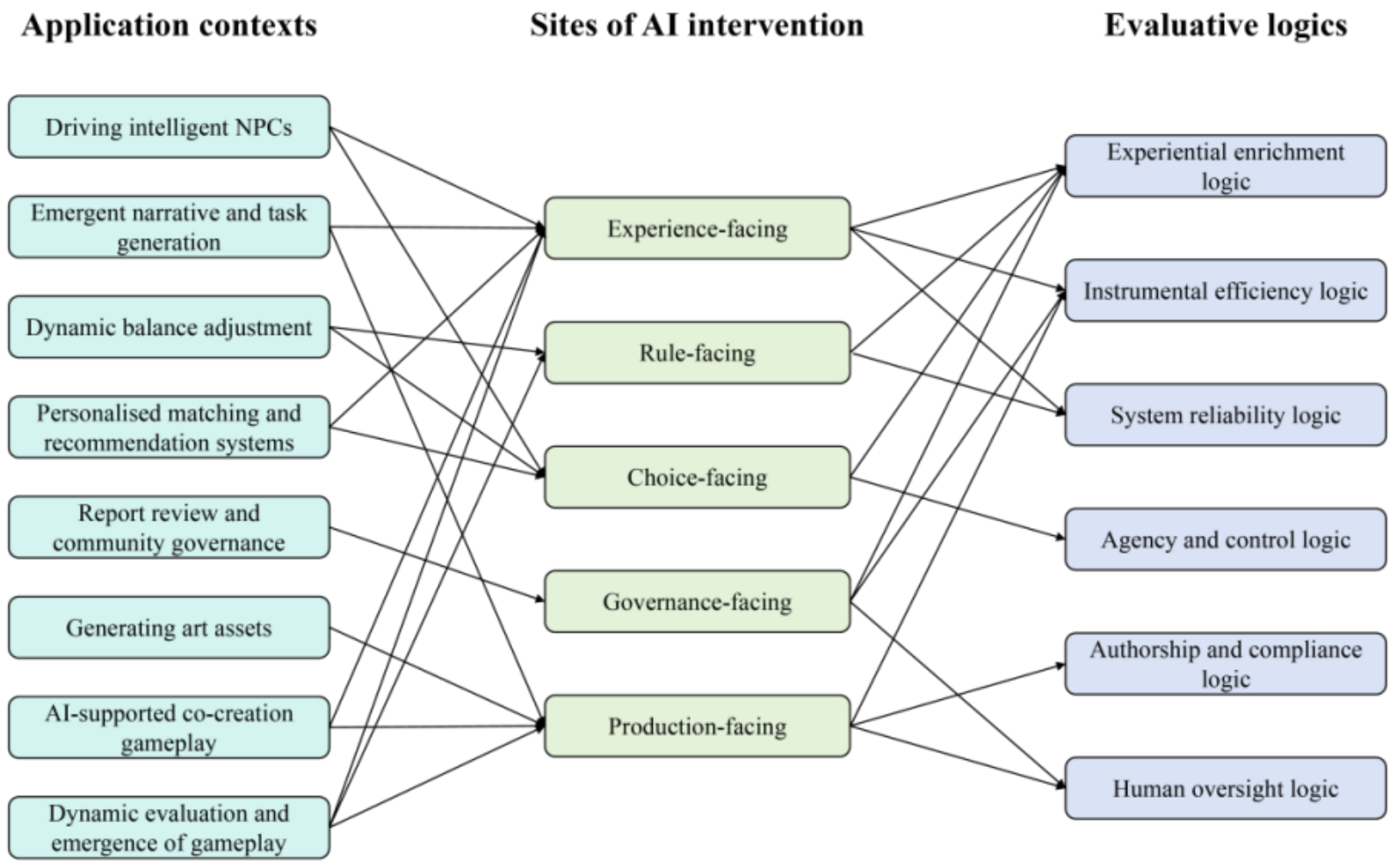

**Fig. 2.** From AI application contexts to evaluative logics: A context-sensitive model of player attitudes

# 4 Discussion

These findings can be interpreted through a context-sensitive view of human-AI interaction. Technology acceptance research emphasizes perceived usefulness as a major predictor of adoption [11], while the machine-agency perspective argues that AI systems are evaluated not only as tools but also as agents that raise concerns about transparency, privacy, and human control [12]. This helps explain why participants did not respond to AI in games as a single category: the same AI capability was acceptable when it expanded valued play experiences, but problematic when it appeared to take over judgment, authorship, or player choice.

In game-specific terms, the six evaluative logics align with an affordance-based understanding of AI user experience, which treats game AI as meaningful insofar as it affords particular forms of action, interpretation, and experience for players [13]. The experiential enrichment logic reflects affordances of presence, personalization, novelty, and emergence. By contrast, the system reliability logic reflects players' concern that AI may damage coherence, balance, rhythm, or learnability. This distinction is consistent with procedural content generation research, which shows that algorithmic generation can support replayability, interactive variation, and design efficiency, but also raises design challenges around control, evaluation, and content quality [14,15].

The agency and human-oversight logics further connect the results to motivation and automation research. Self-determination theory identifies autonomy and competence as central psychological needs [16], which clarifies why dynamic balancing or recommendation could be resisted when it reduced meaningful choice or made achievement feel less earned. Similarly, trust-in-automation research suggests that reliance depends on whether users see automated systems as reliable and appropriately supervised [17]. Participants' preference for AI-assisted but human-led governance and asset creation therefore reflects a theoretically grounded concern with calibrated trust rather than a simple anti-AI position.

Finally, authorship and compliance concerns show that AI-generated game content cannot be discussed only in terms of efficiency or output volume. In creative contexts, players also evaluate whether AI preserves human intentionality, ownership, and accountability. The contribution of this model is both theoretical and practical: it links AI intervention sites to the player values they activate, while offering a diagnostic tool for preserving agency, clarifying oversight, and anticipating resistance in player-centered AI design.

As a work in progress, this study has several limitations. First, open-ended survey responses allowed us to collect diverse player rationales but limited opportunities for deeper probing. Second, descriptive counts were used only to identify salient patterns, and this study does not make statistical claims about population-level prevalence or causal relationships. Third, further analysis is needed to examine whether player background, gaming experience, or genre preferences shape the evaluative logics identified here.

Future work will refine the codebook, further analyze the relationship between attitude polarity, theme combinations, and player background variables, and validate the proposed evaluative-logic model through interviews or scenario-based studies. These steps will help determine whether the evaluative logics identified in this study can generalize beyond the present dataset and inform practical guidelines for player-centered AI design in games.

## 5 Conclusion

This paper examined players' reasons for accepting or rejecting AI across eight application contexts in digital games. The findings show that attitudes are context-dependent and can be summarized through six evaluative logics: experiential enrichment, instrumental efficiency, system reliability, agency and control, authorship and compliance, and human oversight.

By connecting these logics to acceptance, affordance, motivation, and automation theories, the study contributes a context-sensitive framework for explaining player attitudes and guiding AI design in games. Rather than asking only whether AI is acceptable, the framework clarifies where AI intervenes, what player values it supports or threatens, and when human guidance should remain visible. Future work will refine and validate the model through interviews or scenario-based studies.

**Acknowledgments.** Ting-Chen Hsu was responsible for the experimental design, manuscript writing, and data analysis. Jiangxu Lin, Wenran Chen, and Fei Qin jointly conducted the experiments and data coding, sharing the role of second author. The authors thank the participants for their patient responses, which provided indispensable support for the successful completion of this work. This work received no funding.

**Disclosure of Interests.** The authors declare no conflicts of interest.

## Appendix A Questionnaire Design

(1) What is your nickname? (Fill in the blank)

______

(2) What is your age? (Fill in the blank, an integer between 0 and 100)

______

(3) What is your gender?

A. Female B. Male C. Non-binary gender D. Cannot disclose

(4) Which of the following best matches your current academic background?

A. Natural Sciences B. Engineering and Technology C. Humanities and Arts D. Social Sciences E. Interdisciplinary

(5) How much gaming time do you typically play per week?

A. Almost never (less than 1 hour) B. Occasionally (1-5 hours) C. Frequently (6-10 hours) D. Plays a lot (11-20 hours) E. Heavy gamer (more than 20 hours)

(6) What is your opinion on using AI in the following aspects of games?

A. Driving the dialogue, decision-making, and behavior of intelligent NPCs (Opinion____)
B. Emergent generation of narratives or tasks (Opinion ____)
C. Dynamic game balance adjustments (Opinion ____)
D. Personalized matchmaking and recommendation systems (Opinion ____)
E. In-game content review and community governance (Opinion ____)
F. Using AI tools to complete in-game art, voice acting, or text assets (Opinion ____)
G. AI-enabled content co-creation (UGC) gameplay (Opinion ____)
H. Dynamic evolution and emergent generation of gameplay and elements (Opinion ____)

## Appendix B Codebook

Open-ended responses for each scenario must be categorized within the thematic dictionary specific to that scenario. Multiple themes may be extracted from a single response. A single response may simultaneously contain supportive, opposing, and neutral themes. Theme labels should convey a clear orientation (e.g., "Enhances immersion" rather than simply "Immersion").
Scenario 1: Attitudes toward AI Driving Intelligent NPCs [9 themes]
Themes:
[Enhancing immersion/presence]; [High personalization/richness]; [High logicality]; [High freedom]; [Reducing development costs]; [Logical confusion]; [Lack of emotion]; [Out of character]; [Low system stability]
Theme Explanations:
[1] Enhancing immersion/presence: Players believe AI-NPCs can effectively enhance the gaming experience, make characters feel more realistic, increase engagement and investment in the game, and make players more willing to participate.
[2] High personalization/richness: Players believe AI-NPCs can create personalized experiences, enrich game content, make experiences more diverse, and bring greater freshness.
[3] High logicality: Players believe AI-NPCs produce replies with strong logic, high quality, and clear organization.
[4] High freedom: Players believe AI-NPCs significantly increase freedom and openness in the game, allowing players to express ideas more freely.
[5] Reducing development costs: Players believe that developing AI-driven NPCs can effectively reduce workload for developers, improve development efficiency, reduce effort spent on creating NPC settings, and decrease the need for writing dialogue trees.
[6] Logical confusion: Players believe AI-NPC replies often show illogical connections, inconsistencies, irrelevant answers, or nonsensical content.
[7] Lack of emotion: Players believe AI-NPCs lack emotional depth, feel robotic or artificial, become boring over time, and make it difficult to empathize.
[8] Out of character: Players believe AI-NPCs easily go out of character (OOC), show inconsistent personality or reply style, or have fragmented continuity in responses.
[9] Low system stability: Players believe adding AI-NPCs may increase overall game system instability, such as causing bugs, progression blocks, unresponsive NPCs, deviation from main quests, or other system disruptions.
Scenario 2: Attitudes toward AI for Dynamic/Emergent Narrative and Task Generation [8 themes]
Themes:
[Lack of creativity/emotion]; [Low system stability]; [Enhancing immersion]; [High personalization/richness]; [High replayability]; [Chaotic/poor-quality narrative]; [Strong novelty/freshness]; [Extending game lifespan]
Theme Explanations:
[1] Lack of creativity/emotion: Players feel AI-generated narratives lack human creative input, originality, or emotional depth; stories feel formulaic or players simply resist fully AI-generated plots.
[2] Low system stability: Players worry AI narrative generation may cause bugs, progression blocks, or task errors.
[3] Enhancing immersion: Players believe AI-generated narratives effectively deepen game experience and engagement.
[4] High personalization/richness: Players believe AI can meet various narrative preferences, increase content richness and variety, and offer more branches or possibilities.

[5] High replayability: Players believe it improves gameplay value, increases replayability, and reduces repetition.
[6] Chaotic/poor-quality narrative: Emphasizes issues with internal logic, deviation from the main storyline or world-setting, and overall low narrative quality (distinct from system stability issues).
[7] Strong novelty/freshness: Players expect greater freshness, higher anticipation, stronger exploration desire, varied experiences per playthrough, lower repetition, more randomness and surprise, higher openness, and more diverse endings.
[8] Extending game lifespan: Players explicitly mention that AI-generated narratives make the game more enduring and extend its lifecycle.
Scenario 3: Attitudes toward AI for Dynamic Game Difficulty/Balance Adjustment [9 themes]
Themes:
[Enhanced personalization/adaptability]; [Reducing balancing costs]; [Unstable balancing]; [Disrupting game rhythm]; [Enhancing immersion]; [Impaired autonomy]; [Reducing frustration]; [Maintaining challenge]; [Reduced sense of achievement]
Theme Explanations:
[1] Enhanced personalization/adaptability: AI can meet the needs of players at different skill levels and create personalized experiences.
[2] Reducing balancing costs: AI lowers the design effort and repeated manual adjustments required for game balance, improving efficiency.
[3] Unstable balancing: AI adjustments may produce unreasonable or inconsistent balance, inaccuracies, or bugs.
[4] Disrupting game rhythm: AI interference breaks flow state, planned strategies, and overall game feel, turning engaging gameplay monotonous.
[5] Enhancing immersion: AI adjustments improve player engagement and willingness to participate.
[6] Impaired autonomy: Players want to choose their own difficulty, feel their self-efficacy is undermined by "hand-holding," or prefer to attribute difficulty solely to their own skill.
[7] Reducing frustration: Helps lower-skill players avoid frustration and reduces dropout rates (new-player friendly).
[8] Maintaining challenge: Prevents the game from becoming too easy for high-skill players, maintaining motivation through balanced challenge.
[9] Reduced sense of achievement: Dynamic easing makes victories feel unearned or diminishes the reward of overcoming difficulty.
Scenario 4: Attitudes toward AI for Personalized Matching or Recommendation Systems [7 themes]
Themes:
[Prone to monotony]; [Inaccurate matching]; [Enhancing gaming experience]; [Improving gaming efficiency]; [Privacy concerns]; [Catering to diverse preferences]; [Impaired autonomy]
Theme Explanations:
[1] Prone to monotony: Long-term recommendations based on past behavior create filter bubbles, aesthetic fatigue, and lack of novelty.
[2] Inaccurate matching: AI fails to accurately understand player preferences, abilities, or characteristics.
[3] Enhancing gaming experience: Improves immersion, engagement, interaction quality, balance, fairness, etc.
[4] Improving gaming efficiency: Saves time and mental effort in selecting content or opponents.
[5] Privacy concerns: Fear of personal data collection and potential leaks.

[6] Catering to diverse preferences: Effectively meets varied player needs and delivers unique personalized experiences.
[7] Impaired autonomy: Players resist being overly guided or having choices made for them.
Scenario 5: Attitudes toward AI for Report Review or Community Governance [7 themes]
Themes:
[Lack of emotional feedback]; [Strong fairness/objectivity]; [Missed/misjudgment concerns]; [Bias/fairness concerns]; [High efficiency]; [Need for human-AI collaboration]; [Lack of semantic/emotional understanding]
Theme Explanations:
[1] Lack of emotional feedback: AI responses feel robotic, insincere, or perfunctory.
[2] Strong fairness/objectivity: AI delivers logical, consistent, and fair rulings in most cases.
[3] Missed/misjudgment concerns: AI may be unstable, miss violations, or wrongly classify content.
[4] Bias/fairness concerns: AI may systematically favor certain player groups.
[5] High efficiency: Much faster processing and timely feedback than human moderation.
[6] Need for human-AI collaboration: Players prefer a hybrid approach rather than full AI or full human control.
[7] Lack of semantic/emotional understanding: AI struggles with complex human behavior, sarcasm, slang, or nuanced social contexts.
Scenario 6: Attitudes toward AI for Art Asset Creation [12 themes]
Themes:
[AI-assisted & human-led]; [Copyright risks]; [Weak innovation/creativity]; [Low aesthetics]; [High aesthetics]; [Poor controllability]; [Benefiting developers]; [Lack of creator empathy]; [Low generation quality]; [High generation quality]; [Lack of authenticity/emotion]; [Strong perfunctory feeling]
Theme Explanations:
[1] AI-assisted & human-led: AI can provide references or inspiration, but core creative work should remain human-driven.
[2] Copyright risks: Concerns about infringement or unclear legal status of AI-generated assets.
[3] Weak innovation/creativity: Outputs feel generic, "AI-like," and lack the spark of human creativity.
[4] Low aesthetics: Reduces visual/auditory quality and causes aesthetic downgrade.
[5] High aesthetics: Delivers superior visual/auditory quality compared to traditional methods.
[6] Poor controllability: Frequent hallucinations, anatomical errors, style inconsistency, etc.
[7] Benefiting developers: Accelerates iteration, lowers costs, and improves development efficiency.
[8] Lack of creator empathy: Players cannot emotionally connect with or feel the human artistic intent behind the work.
[9] Low generation quality: Assets look bad, sound bad, or feel uninteresting.
[10] High generation quality: Assets look good, sound good, and feel high-quality.
[11] Lack of authenticity/emotion: Images feel unreal, music/voice lacks soul, etc.
[12] Strong perfunctory feeling: Use of AI assets feels disrespectful to players, as if the developers cut corners.
Scenario 7: Attitudes toward AI Supporting Player Co-Creation of Personalized Content [11 themes]
Themes:
[Copyright/compliance concerns]; [Lowering creation threshold]; [Meeting diverse demands]; [Improving creation efficiency]; [Enhancing immersion]; [Enhancing novelty/richness]; [Quality concerns]; [High freedom/openness]; [Lack of authorship]; [Meeting sense of achievement]; [Meeting sharing desire]
Theme Explanations:
[1] Copyright/compliance concerns: Risk of infringement, violation of platform rules, ethics, or law.
[2] Lowering creation threshold: Enables players with limited skills to realize their ideas.
[3] Meeting diverse demands: Allows realization of varied personal visions and unique experiences.
[4] Improving creation efficiency: Saves time realizing ideas.
[5] Enhancing immersion: Increases engagement and investment in the game.
[6] Enhancing novelty/richness: Continuously introduces new, unique elements.
[7] Quality concerns: Generated content may be low-quality or fail to meet expectations.
[8] High freedom/openness: Expands game boundaries and player creative freedom.
[9] Lack of authorship: Players feel the creations are not truly "theirs."
[10] Meeting sense of achievement: The creation process provides strong accomplishment.
[11] Meeting sharing desire: Encourages players to share their creations.
Scenario 8: Attitudes toward AI for Emergent/Dynamic Gameplay Mechanisms [8 themes]
Themes:
[Meeting diverse demands]; [Content quality concerns]; [Disrupting balance/system stability]; [Improving development efficiency]; [Increasing cognitive load]; [Enhancing immersion/exploration desire]; [Increasing novelty/richness]; [Extending game lifespan]
Theme Explanations:
[1] Meeting diverse demands: Realizes many player-desired mechanics and personalized experiences.
[2] Content quality concerns: Emergent elements may be low-quality or poorly implemented.
[3] Disrupting balance/system stability: New elements break designed balance, world tone, or cause bugs and chaos.
[4] Improving development efficiency: Reduces developer workload for new content, mods, or expansions.
[5] Increasing cognitive load: Constant new elements cause fatigue and make deep system mastery difficult.
[6] Enhancing immersion/exploration desire: Encourages deeper engagement and motivation to explore.
[7] Increasing novelty/richness: Provides ongoing fresh experiences and greater content depth.
[8] Extending game lifespan: Makes the game more enduring and worthy of long-term play.